\documentstyle[graphicx,prd,aps]{revtex}
\begin{document}

\twocolumn[                              
\hsize\textwidth\columnwidth\hsize\csname@twocolumnfalse\endcsname  

\title{Cosmic Solenoids:\\
Minimal Cross-Section and Generalized Flux Quantization}
\author{Aharon Davidson and David Karasik}
\address{Physics Department, Ben-Gurion University of the
Negev, Beer-Sheva 84105, Israel \\
(davidson@bgumail.bgu.ac.il, karasik@bgumail.bgu.ac.il)}
\maketitle

\begin{abstract}
A self-consistent general relativistic configuration
describing a finite cross-section magnetic flux tube is
constructed.
The cosmic solenoid is modeled by an elastic superconductive
surface which separates the Melvin core from the surrounding
flat conic structure.
We show that a given amount $\Phi$ of magnetic flux cannot be
confined within a cosmic solenoid of circumferential radius
smaller than ${\displaystyle \frac{\sqrt{3G}}{2\pi c^{2}}\Phi}$
without creating a conic singularity (the expression for the
angular deficit is different from that naively expected).
Gauss-Codazzi matching conditions are derived by means of a
self-consistent action.
The source term, representing the surface currents, is
sandwiched between internal and external gravitational
surface terms.
Surface superconductivity is realized by means of a Higgs
scalar minimally coupled to projective electromagnetism.
Trading the 'magnetic' London phase for a dual 'electric'
surface vector potential, the generalized quantization
condition reads:
${\displaystyle \frac{e}{hc}\Phi+\frac{1}{e}Q=n}$ with $Q$
denoting some dual 'electric' charge, thereby allowing for
a non-trivial Aharonov-Bohm effect.
Our conclusions persist for dilaton gravity provided the
dilaton coupling is sub-critical.
\end{abstract}
\pacs{PACS numbers:}]

\section{Introduction}
The Dirac procedure of squeezing a magnetic flux tube,
while keeping its total magnetic flux $\Phi$ fixed, plays
an important role in theoretical physics.
It is usually assumed that one can shrink the solenoid into
a thin magnetic flux string characterized by the potential
\begin{equation}
	A_{\mu}dx^{\mu} =
	Nd\phi = \frac{N}{r^{2}}(xdy-ydx) ~.
	\label{flux}
\end{equation}
Such a measure zero infinitely-long flux string is of course
classically invisible, but can still allow for a non-trivial
quantum mechanical Aharonov-Bohm effect\cite{aharonov}.
Furthermore, if the total magnetic flux is properly quantized,
namely
\begin{equation}
	\frac{e\Phi}{hc} = n ~,
	\label{n}
\end{equation}
the flux string can be regarded an artifact since it cannot
be detected by test particles carrying quantized electric
charge in units of $e$.
In which case, eq.(\ref{flux}) is nothing but a pure gauge.
By the same token, the semi-infinite magnetic flux string
attached to a Dirac magnetic monopole\cite{dirac}, carrying
magnetic charge $g$, becomes physically irrelevant in case
that
\begin{equation}
	\frac{eg}{hc} = \frac{1}{2}n ~.
\end{equation}

On dimensional grounds, however, it is clear that a string
is not really capable of fully representing the case of an
arbitrarily narrow tube.
In particular, the innocent looking magnetic flux string
configuration eq.(\ref{flux}) is apparently sourceless.
To keep track of the source, one should consider a magnetic
flux tube of finite radius, and let a surface current constitute
the source.
This way, to keep the total magnetic flux finite, the surface
current gets infinitely large as the tube becomes infinitesimally
narrow.

The situation is conceptually and drastically changed once gravity
(or string theory) enters the game.
In this paper, a self-consistent general relativistic configuration
describing a finite cross-section cosmic solenoid is constructed.
The cosmic solenoid is modeled by an elastic superconductive
surface which separates the curved inner core from the surrounding
flat conic structure.
Any attempt to squeeze the cosmic solenoid, while holding
its total magnetic flux $\Phi$ fixed, comes with a cosmic penalty.
In the extreme, if trying to imprison any given amount of magnetic
flux within a tube of a sub-critical cross-section, one pays the
ultimate price of closing the surrounding space and creating a
conic singularity.

\smallskip
Adopting the $\hbar = c =1$ units, here are some exact spacetime
configurations\cite{exact} relevant to our discussion:

\smallskip
\noindent $\bullet$ The general \textit{stationary} cylindrically
symmetric \textit{vacuum} solution of Einstein equations is given
by
\begin{equation}
	ds^{2}=-\Omega (dt+Vd\phi)^{2}+\omega (dr^{2}+
	dz^{2})+\frac{r^{2}}{\Omega}d\phi^{2} ~,
\end{equation}
where we have used the notations
\begin{equation}
	\begin{array}{rcl}
		\omega (r) &=& \displaystyle{r^{\frac{1}{2}(n^{2}-1)}} ~,  
		\vspace{6pt} \\
		\Omega (r) &=& \displaystyle{r\left(\frac{\alpha}{r^{n}} -
		\beta r^{n}\right) }~,  \vspace{6pt} \\
		V(r) &=& \displaystyle{\sqrt{\frac{\beta }{\alpha}}
		\frac{r^{n+1}}{\Omega}} ~.		
	\end{array}
\end{equation}

\smallskip
\noindent $\bullet$ The \textit{static} solution calls for
$V(r)=const$, and can be put in the familiar $(a,b,c)$ Kasner
form
\begin{equation}
	ds^{2}= -r^{2a}dt^{2} + r^{2b}dz^{2} + dr^{2}+
	\gamma^{2}r^{2c}d\phi^{2} ~,
	\label{kasner}
\end{equation}
with the various parameters subject to
\begin{equation}
	\begin{array}{c}
		a+b+c=1 ~,  \vspace{6pt}\\
		a^{2}+b^{2}+c^{2}=1 ~.
	\end{array}
\end{equation}
The factor $\gamma$ acquires a (global) physical meaning once
the periodicity notation is specified, say
\begin{equation}
	\Delta\phi=2\pi  ~.
\end{equation}

The only asymptotically-flat Kasner solution is the locally-flat
yet globally conic $(0,0,1)$ solution.
The conic defect, reflecting the topological charge of the
sectional $2$-metric, measures the variation of the invariant
circumferential radius $\rho(r)$ with respect to the invariant
radial distance $R(r)$.
Given the flat Kasner metric eq.(\ref{kasner}), for which
$\rho=\gamma r$ and $R=r$, one finds
\begin{equation}
	\frac{d\rho}{dR}=\gamma ~,
\end{equation}
corresponding to a deficit angle of $2\pi (1-\gamma)$.
As long as $\gamma \geq 0$, the outer space is open, but
$\gamma<0$ takes us to the back side of the cone.
In which case, the outer space is closed and exhibits a
singular point at a finite distance. 

A well known example is the space surrounding a cosmic string
\cite{Vilenkin,strings,SCstrings}.
The corresponding Vilenkin conic defect\cite{Vilenkin} reads
\begin{equation}
	\gamma = 1-4\mu G ~,
	\label{gammas}
\end{equation}
where $\mu$ is the energy density per unit length of the string.
The physical demand $\gamma \geq 0$ then leads to the consistency
condition $\displaystyle{\mu \leq \frac{1}{4G}}$.

\smallskip
\noindent $\bullet$ The general cylindrically-symmetric
\textit{static} solution of Einstein-Maxwell equations involving
a longitudinal magnetic field (caused by an angular current) is
referred to as the Witten solution\cite{Witten}
\begin{equation}
	ds^{2}=\omega^{4}W^{2}(-dt^{2}+dr^{2})+W^{2}dz^{2}+
	\frac{r^{2}}{W^{2}}d\phi^{2} ~,
\end{equation}
where the Witten factor $W(r)$ is given by
\begin{equation}
	W(r)=\frac{\alpha}{r^{n-1}}+\beta r^{n+1} ~.
\end{equation}
This metric can be viewed as a soliton connecting two Kasner
regimes, namely $\left(\frac{n(n-1)}{n^{2}-n+1},\frac{1-n}
{n^{2}-n+1},\frac{n}{n^{2}-n+1}\right)$ and its $n\rightarrow-n$
companion.

\smallskip
\noindent $\bullet$ Regularity at the $r=0$ axis of symmetry (at
the expense of a generic singularity at $r\rightarrow\infty$) singles
out the Lorentz invariant (along the $z$-axis) $\alpha^{2}=n=1$ case,
known as the Melvin solution\cite{Melvin}
\begin{equation}
	ds^{2} = M^{2}(r)(-dt^{2}+dz^{2}+dr^{2})+
	\frac{r^{2}}{M^{2}(r)}d\phi^{2} ~,
	\label{melvin}
\end{equation}
where the Melvin factor $M(r)$ is given by
\begin{equation}
	M(r) \equiv 1+\frac{1}{4}GB^{2}r^{2} ~.
\end{equation}
It is accompanied by the electromagnetic configuration
\begin{equation}
	\begin{array}{rcl}
		A_{\phi}& = & \displaystyle{\frac{Br^{2}}{2M(r)}} ~, 	 
		\vspace{6pt} \\
		F_{r\phi} & = & \displaystyle{\frac{Br }{M^{2}(r)}} ~.
	\end{array}
\end{equation}
One observes that the magnetic flux is practically confined within a
tube of radius $\displaystyle{\frac{2}{\sqrt G}B}$, where $B$ is the
value of the magnetic field at the origin (in our form notations,
$\displaystyle{B_{z}=\frac{1}{r}F_{r\phi}}$).
As expected, the Melvin solution represents a soliton which connects
the only two Kasner branches which are allowed by partial Lorentz
invariance, namely
\begin{equation}
	(0,0,1) \longleftrightarrow
	(\frac{2}{3},\frac{2}{3},-\frac{1}{3}) ~.
\end{equation}
Whereas the $(0,0,1)$ branch, associated with $r \rightarrow 0$, is
regular, the $(\frac{2}{3},\frac{2}{3},-\frac{1}{3})$ branch, associated
with $r\rightarrow\infty$, is not only singular but furthermore exhibits
a vanishing circumferential radius.
\begin{figure}[tbp]
	\begin{center}
		\includegraphics[scale=0.7]{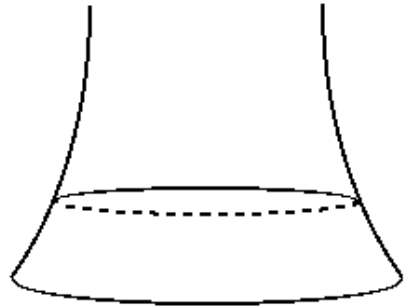}
		\caption{Melvin bone: The 2-dim surface (as embedded in
		flat 3-space) characterized by the sectional 2-metric
		$ds^{2}=M^{2}(r)dr^{2}+r^{2}M^{-2}(r)d\phi^{2}$.}
		\label{fig1}
	\end{center}
\end{figure}
In light of this, a cosmic solenoid should be constructed by pasting
together two manifolds.
The core of the tube, which contains the magnetic flux and consequently
exhibits the Melvin geometry, is wrapped by a flat Kasner Universe
which carries no magnetic fields but has a generic conic structure.
These manifolds are separated by an elastic superconductive surface
hosting the surface currents.
In a self-consistent model like ours, these surface currents are
expected to be dynamically created by some surface fields minimally
coupled to projective electromagnetism. 
The corresponding self-consistent action principle must not only
govern the equations of motion, but also the matching conditions
(including in particular the Gauss-Codazzi formalism) of the various
fields involved.
The latter include the gravitational field $g_{\mu\nu}$, the
electromagnetic vector potential $A_{\mu}$, the optional dilaton
field $\eta$, and a variety of $(2+1)$-dimensional surface fields.
It is only after such a model is analytically constructed, that one
is able to analyze the case from the shrinking circumferential radius
point of view.

This paper is organized as follows.
First, we present the tenable action principle and derive
the corresponding equations of motion and the attached matching
conditions.
In this Lagrangian formalism, the surface source term is
sandwiched between the internal and the external gravitational
surface terms. 
This is the prescription behind the field theoretical recovery
of the Gauss-Codazzi and the electromagnetic matching equations.
We are after the cylindrically symmetric static solution,
focusing our attention on two parameters, namely the
circumferential radius of the tube and the conic defect of
the surrounding space-time.
Surface superconductivity is realized by means of a Higgs
scalar minimally coupled to projective electromagnetism,
thereby establishing the linkage between the circumferential
radius and the conic defect. 
The main result of this linkage is that the circumferential
radius $\rho$ cannot be arbitrarily small unless the surrounding
space-time suffers a singularity.
To be more specific (and use momentarily the full units),
\begin{equation}
	\fbox{$\displaystyle{\frac{}{}
	\rho_{min}\geq\frac{\sqrt{3G}}{2\pi c^{2}}\Phi}$}
\end{equation}
Notice that this bound is $\hbar$-independent, indicating that
this result is purely gravitational and has nothing to do with
quantum mechanics.
Trading the 'magnetic' London phase for an 'electric' dual
surface vector field, we then replace eq.(\ref{n}) by the
generalized quantum mechanical quantization condition
\begin{equation}
	\fbox{$\displaystyle{
	\frac{e\Phi}{hc} + \frac{1}{e}Q =n}$}
\end{equation}
where $Q$ is the dual electric charge involved.
Notice the fact that $\Phi$ itself is not quantized, thereby
opening the door for a non-trivial Aharonov-Bohm effect.
Finally, we offer a similar treatment to a cosmic solenoid in
dilaton gravity background, and verify the validity of our main
results as long as the dilaton coupling remains sub-critical.

\section{Self-consistent action principle}
Up to as yet unspecified model-dependent source, the action
principle is quite conventional
\begin{eqnarray}
	&& Action = \nonumber \vspace{6pt} \\
	&& - \frac{1}{16\pi}\int_{V_{in}}
	(g^{\mu\lambda}g^{\nu\sigma}F_{\mu\nu}F_{\lambda\sigma})^{in}
	\sqrt{-\det g}\,d^{4}x + \nonumber \\
	&& + \frac{1}{16\pi G} \int_{V_{in}}{\cal R}^{in}
	\sqrt{-\det g}\,d^{4}x - \nonumber \\
	&& - \frac{1}{8\pi G} \oint_{S} K^{in}
	\sqrt{-\det\gamma}\,d^{3}y + \nonumber \\
	&& + \oint_{S} {\cal L}^{S}\sqrt{-\det\gamma}\,d^{3}y -
	\label{action} \\
	&& - \frac{1}{8\pi G} \oint_{S} K^{out}
	\sqrt{-\det\gamma}\,d^{3}y + \nonumber \\
	&& + \frac{1}{16\pi G} \int_{V_{out}} {\cal R}^{out}
	\sqrt{-\det g}\,d^{4}x - \nonumber \\
	&& - \frac{1}{16\pi}\int_{V_{out}}(g^{\mu\lambda}g^{\nu\sigma}
	F_{\mu\nu}F_{\lambda\sigma})^{out}\sqrt{-\det g}\,d^{4}x ~.
	\nonumber
\end{eqnarray}
The canonical fields are $g_{\mu\nu}(x)$ and $A_{\mu}(x)$, which are
continuous over the surface, and some surface degrees of freedom (soon 
to be specified).
In our notations, 
\begin{equation}
	\gamma_{ab}=g_{\mu\nu}(x(y))x^{\mu}_{\,,a}x^{\nu}_{\,,b}
\end{equation}
is the induced $(2+1)$-dimensional metric,	$K_{\mu\nu}^{in/out}$ is
the extrinsic curvature of the surface $S$ embedded in $V_{in/out}$,
and $K=g^{\mu\nu}K_{\mu\nu}$.
Each gravitational action in $V_{in/out}$ is accompanied by its own
surface term \cite{boundary}.
This way, the surface matter action is sandwiched between the two
gravitational surface terms. 
Obviously, the two gravitational surface terms are expected to cancel
each other in the empty case, where the surface is regarded an artifact.
A similar idea has been recently suggested in the context of black hole 
membranes\cite{sandwich}.

\smallskip
\noindent $\bullet$ The variation of the action with respect to $g_{\mu\nu}(x)$
gives rise to the Einstein equations in the entire spacetime
\begin{equation}
	\left.{\cal R}_{\mu\nu} - \frac{1}{2}g_{\mu\nu}{\cal R} -
	8\pi G T_{\mu\nu}\right|_{in/out}=0 ~,
	\label{Ein}
\end{equation}
and produces the Gauss-Codazzi matching condition on the surface
\begin{equation}
	\left. {\cal K}_{ab} \right|_{in} +
	\left.{\cal K}_{ab}\right|_{out} = 8\pi GS_{ab} ~,
	\label{GC}
\end{equation}
where
\begin{equation}
	{\cal K}	_{ab} \equiv
	(Kh_{\mu\nu}-K_{\mu\nu})x^{\mu}_{,a}x^{\nu}_{,b} ~,
\end{equation}
and $h_{\mu\nu}$ is the first fundamental form of the surface.
The energy-momentum tensor in $V_{in/out}$ is
\begin{equation}
	T_{\mu\nu}^{in/out} = - 2\frac{\partial {\cal L}^{in/out}}
	{\partial g^{\mu\nu}} + {\cal L}^{in/out}g_{\mu\nu} ~.
\end{equation}
In a similar way, the surface energy-momentum tensor is defined
via
\begin{equation}
	S_{ab} = -2\frac{\partial {\cal L}^{S}}
	{\partial \gamma^{ab}} + {\cal L}^{S}\gamma_{ab}~.
\end{equation}
Eq. (\ref{GC}) is the matching equation suggested by Israel
\cite{Israel} (but has not been derived by means of a Lagrangian
formalism).
In its present form it solely involves surface projections, a
property which in turn allows us to use different coordinate
systems inside and outside the tube.

\smallskip
\noindent $\bullet$ The variation of the action with respect to
$A_{\mu}(x)$ gives rise to the Maxwell equations in the entire
spacetime
\begin{equation}
	\left. F^{\mu\nu}_{;\nu}\right|_{in/out} = 0 ~,
	\label{Max}
\end{equation}
and produces the electromagnetic junction conditions on the surface
\begin{equation}
	\left. F_{\mu\nu} n^{\mu} x^{\nu}_{,a}\right|_{in} +
	\left. F_{\mu\nu} n^{\mu} x^{\nu}_{,a}\right|_{out} =
	4\pi \gamma_{ab}j^{b} ~,
	\label{Ms}
\end{equation}
where $n_{\mu}^{in/out}$ is the unit outside/inside-pointing normal vector.
The surface current is given by
\begin{equation}
	j^{a} = \frac{\partial {\cal L}^{S}}{\partial A_{a}} ~,
	\label{ja}
\end{equation}
with $A_{a}=A_{\mu}x^{\mu}_{,a}$ denoting the projective electromagnetic
vector potential.

\section{Cosmic solenoid}
The cosmic solenoid solution is subject to the following requirements:
\begin{enumerate}
	\item The configuration is static, cylindrically symmetric, and
	partially Lorentz invariant. It admits three Killing vectors:
		${\displaystyle \frac{\partial}{\partial t}}$,
		${\displaystyle \frac{\partial}{\partial z}}$,
		${\displaystyle \frac{\partial}{\partial \phi}}$.
	\item The surface is stable, defined by $r=R=\text{const}$.
	\item The inner space is conic singularity free.
	\item The outer spacetime is asymptotically flat and
		free of electromagnetic fields.
\end{enumerate}
Solving Einstein-Maxwell equations, given the above constraints,
one can immediately verify that

\medskip
\noindent $\bullet$ The \textit{inner} solution is the Melvin
Universe
\begin{mathletters}
\label{melvinin}
	\begin{equation}
		ds^{2}_{in} =
		M^{2}(r)(-dt^{2}+dz^{2}+dr^{2})+
		\frac{r^{2}}{M^{2}(r)}d\phi^{2} ~,
	\end{equation}
	\begin{equation}
		F_{r\phi}^{in} = \frac{Br}{M^{2}(r)} ~,~~
		A_{\phi}^{in} = \frac{Br^{2}}{2M(r)} ~.
	\end{equation}
\end{mathletters}
Notice that, as $r\rightarrow 0$, $A_{\phi}^{in}$ vanishes (no 
residual flux strings) and $ds^{2}_{in}$ approaches Minkowski
line element.

\medskip
\noindent $\bullet$ The \textit{outer} solution is the (0,0,1)
Kasner Universe with a conic structure
\begin{mathletters}
\label{kasnerin}
	\begin{equation}
		ds^{2}_{out} =
		-dt^{2}+dz^{2}+dr^{2}+\gamma^{2}r^{2}d\phi ^{2}~,
	\end{equation}
	\begin{equation}
		F_{r\phi}^{out} = 0 ~, \quad A_{\phi}^{out} = N ~.
	\end{equation}
\end{mathletters}
The total magnetic flux confined in the tube is
\begin{equation}
	\Phi=\oint A_{\phi}d\phi=2\pi N ~.
\end{equation}

The surface equation is obtained by the embedding $x^{\mu}_{in/out}(y)$.
Each of the metrics (\ref{melvinin}) and (\ref{kasnerin}) was written in a
convenient coordinate system; one must thus use different coordinate systems
inside and outside the tube.
Both line elements exhibit Lorentz invariance in the $z$
direction, and $\Delta \phi=2\pi$ periodicity.
Therefore, the most general coordinate transformation between
the inner system and the outer system is given by
\begin{eqnarray}
	t^{out}= & t^{S}= & \lambda t^{in} ~. \nonumber \\
	z^{out}= & z^{S}= & \lambda z^{in} ~. \\
	\phi^{out}= & \phi^{S}= & \phi^{in} ~. \nonumber
\end{eqnarray}
The location of the separating surface is defined by both
\begin{equation}
	\begin{array}{c}
		r^{in} =  R^{in} ~, \vspace{6pt}	\\
		r^{out}=  R^{out} ~,	
	\end{array}
\end{equation}
but there is no reason why should $R^{in/out}$ be taken equal.

Substitute Eqs. (\ref{melvinin},\ref{kasnerin}) in Eqs. 
(\ref{Ein}-\ref{ja}), one derives the following matching relations:

\smallskip
\noindent $\bullet$ Keeping the metric continuous over the surface
determines the coordinate transformation and the conic defect,
namely
\begin{equation}
	\lambda = 1 + \frac{1}{4}GB^{2}R_{in}^{2} ~,
\end{equation}
\begin{equation}
	\gamma = \frac{R_{in}}{R_{out}
	\left(1+\frac{1}{4}GB^{2}R_{in}^{2}\right)} ~.
\end{equation}

\smallskip
\noindent $\bullet$ Calculate the extrinsic curvature of the surface,
with respect to the inside/outside embeddings, and use eq.(\ref{GC})
to obtain
\begin{equation}
	S^{t}_{t} = S^{z}_{z} = \frac{1}{8\pi G}\left( \frac{1}{R_{out}} -
	\frac{1}{R_{in}\left(1+\frac{1}{4}GB^{2}R_{in}^{2}\right)}\right) ~,
\end{equation}
\begin{equation}
	S^{\phi}_{\phi} = - \frac{B^{2}R_{in}}
	{8\pi\left(1+\frac{1}{4}GB^{2}R_{in}^{2}\right)^{2}} ~.
\end{equation}

\smallskip
\noindent $\bullet$ The continuity of the electromagnetic vector potential
across the surface leads to:
\begin{equation}
	N = \frac{BR_{in}^{2}}{2\left(1+\frac{1}{4}GB^{2}R_{in}^{2}\right)} ~.
\end{equation}

\smallskip
\noindent $\bullet$ The junction condition for the electromagnetic 
fields reads
\begin{equation}
	4\pi j^{\phi} =
	\frac{B}{R_{in}\left(1+\frac{1}{4}GB^{2}R_{in}^{2}\right)} ~.
\end{equation}

The set of matching equations tells us that the yet unspecified surface
fields must fulfill two consistency conditions, namely
\begin{mathletters}
\label{consist}
	\begin{equation}
        	S^{t}_{t}=S^{z}_{z}=-\sigma ~, 
		\label{Edensity}
	\end{equation}
\begin{equation}
	S^{\phi}_{\phi}+Nj^{\phi}=0 ~.
	\label{tension}
\end{equation}
\end{mathletters}
Eq.(\ref{Edensity}) reflects the Lorentz invariance in the $z$ direction,
and defines the surface energy density $\sigma$.
Eq. (\ref{tension}) expresses the equilibrium between the gravitational
attraction and the electromagnetic repulsion.

An observer living in the outer spacetime is actually aware of  only
three parameters. 
They are: 
\begin{enumerate}
	\item  The total flux confined in the tube $\Phi=2\pi N$,
	\item  The conic defect $\gamma$, which can be measured by 
	gravitational lensing, and
	\item  The circumferential radius $\rho$ of the tube, given by
	the double relation $\displaystyle{\rho=\frac{R_{in}}{1+\frac{1}{4}
	GB^{2}R_{in}^{2}} = \gamma R_{out}}$.
\end{enumerate}
Holding the total magnetic flux $2\pi N$ fixed, in accord with
Dirac procedure, the circumferential radius of the tube gets
related to the surface current
\begin{equation}
	\left( \rho + G\frac{N^{2}}{\rho} \right)^{3} =
	\frac{N}{2\pi j^{\phi}} ~,
	\label{circ}
\end{equation}
and the conic defect
\begin{equation}
	\gamma = \frac{1}{\displaystyle{
	\left(1+G\frac{N^{2}}{\rho^{2}}\right)^{2}}} - 4\mu_{s} G  
\end{equation}
gets related to $\mu_{s}$, the \textit{surface} energy per unit
length
\begin{equation}
	\mu_{s}=2\pi\rho\sigma ~.
\end{equation}
The quantity $\mu_{s}$ need not be confused with $\mu_{in}$,
the energy per unit length \textit{in} the tube.
The latter can be calculated by integrating the $T^{0}_{0}$
component of the energy momentum tensor over a plane perpendicular
to the solenoid, namely
\begin{equation}
	\mu_{in}=-\int_{0}^{R_{in}}M(r)dr
	\int_{0}^{2\pi}\frac{rd\phi}{M(r)}T^{0}_{0} ~.
\end{equation}
Substituting
\begin{equation}
	T^{0}_{0}=\frac{1}{4\pi}F^{0\mu}F_{0\mu}-
	\frac{1}{16\pi}\delta^{0}_{0}F^{\mu\nu}F_{\mu\nu}=
	-\frac{1}{8\pi}\frac{B^{2}}{M^{4}(r)} ~,
\end{equation}
with $M(r)$ denoting the Melvin factor, we find
\begin{equation}
	\mu_{in}=\frac{1}{6G}\left[1-
	\frac{1}{\left(\displaystyle{1+G\frac{N^{2}}
	{\rho^{2}}}\right)^{3}}\right] ~.
\end{equation}
In turn, the conic defect takes the final form
\begin{equation}
	\fbox{$\displaystyle{
	\gamma=(1-6\mu_{in}G)^{2/3}-4\mu_{s}G}$}
	\label{gamma}
\end{equation}
thereby constituting the flux-tube generalization of the cosmic
string conic defect eq.(\ref{gammas}).
For large $\rho$, that is in the Nambu-Goto limit, we do recover
\begin{equation}
	\gamma \approx 1-4(\mu_{in}+\mu_{s})G ~.
\end{equation}

\section{Surface superconductivity}
The missing pieces of the puzzle are of course the $(2+1)$-dimensional
surface fields.
By virtue of their coupling to the $(3+1)$-dimensional fields, they
serve as sources. 
And by exhibiting their own equation of motion, whose solution
must obey eqs.(\ref{consist}), they make self-consistent sources.
For simplicity, we consider first the prototype case of a complex
scalar field\cite{phiasQ} minimally coupled to projective
electromagnetism, and then discuss the dual case.

\subsection{Complex scalar field}
The simplest source term
\begin{equation}
	{\cal L}^{S}= -\gamma^{ab}(D_{a}\Psi)^{\dagger}(D_{b}\Psi)
	-V(\Psi^{\dagger}\Psi) ~,
\end{equation}
involves a complex scalar field $\Psi(y)=\xi (y)e^{i\chi(y)}$ minimally
coupled to the projective electromagnetic field.
This is realized by means of the covariant derivative
\begin{equation}
	D_{a}\Psi = (\frac{\partial}{\partial y^{a}}-ieA_{a})\Psi ~,
\end{equation}
where $A_{a} = A_{\mu}(x)x^{\mu}_{,a}$ and dimensionless $e$
being the electromagnetic coupling constant.
Invoking the gauge invariant quantity
\begin{equation}
	\Delta_{a}\equiv\chi_{,a}-eA_{a} ~,
\end{equation}
the equations of motion are given by
\begin{mathletters}
\label{scalar}
\begin{equation}
	(\gamma^{ab}\xi_{;a})_{;b}-
	\xi(\gamma^{ab}\Delta_{a}\Delta_{a}+
	\frac{dV}{d\xi^{2}}) = 0 ~,
\end{equation}
\begin{equation}
        j^{a}_{;a} \equiv
	\left( 2e\xi^{2}\gamma^{ab}\Delta_{b}\right)_{;a} = 0 ~.
\end{equation}
\end{mathletters}
The only static solution of these equations which does not upset
the consistency conditions eqs.(\ref{consist}) is of the London
type, that is
\begin{equation}
	\left\{
	\begin{array}{rcl}
		\xi & = & m = const ~,  \\
		\chi & = & n\phi ~,
	\end{array}
	\right.
\end{equation}
with integer $n$ keeping the extremal field configuration
single-valued.
The fact that $m\neq 0$ means that $U(1)_{EM}$ is spontaneously 
violated on (and only on) the tube surface.
This is the group theoretical origin of the superconductive
currents, the generators of the imprisoned magnetic flux.

A crucial role is played here by the effective potential
\begin{equation}
	V_{eff} = V + \xi^{2}\gamma^{ab}\Delta_{a}\Delta_{a} ~,
\end{equation}
having the properties that
\begin{equation}
	\left\{
	\begin{array}{rcl}
		{\displaystyle \left. V_{eff}\right|_{\xi= m}}
		& = & \sigma ~, \vspace{6pt} \\
		{\displaystyle \left.\frac{dV_{eff}}{d\xi^{2}}
		\right|_{\xi= m}} & = & 0 ~.
	\end{array}
	\right.
\end{equation}
Using the definition of $S^{a}_{\,b}$ and the consistency
relations eq.(\ref{consist}), we infer that
\begin{equation}
	\sigma = 2\frac{m^{2}}{\rho^{2}}n(n-eN) ~.
\end{equation}
In turn, we can write
\begin{equation}
	\left\{
	\begin{array}{rcl}
		{\displaystyle \left. V\right|_{\xi= m}}
		\equiv \Lambda
		& = & {\displaystyle \frac{m^{2}}{\rho^{2}}
		\left(n^{2} - e^{2}N^{2}\right)} ~, \vspace{6pt} \\
		{\displaystyle \left.\frac{dV}{d\xi^{2}}
		\right|_{\xi= m}}
		& = &{\displaystyle -\frac{(n-eN)^{2}}{\rho^{2}}} ~.
	\end{array}
	\right.
	\label{potential}
\end{equation}
The insertion of a surface field establishes the desired link
between the surface current $j^{a}$ and the surface energy
density $\sigma$
\begin{equation}
	\sigma = \frac{n}{e}j^{\phi} = \frac{2n\Lambda}{n+eN} ~.
	\label{n/e}
\end{equation}

We now claim that the physical solution dictates
\begin{equation}
	\left\{
	\begin{array}{rcl}
		&\Lambda >0 ~,& \vspace{6pt}\\ 
		&{\displaystyle \frac{n}{eN}>1} ~.&
	\end{array}
	\label{eN/n}
	\right.
\end{equation}
To see the point, recall the two relations
\begin{mathletters}
	\begin{equation}
		e^{2}N^{2}+\frac{\rho^{2}}{m^{2}}
		\Lambda=n^{2} ~,
	\label{quan1}
\end{equation}
\begin{equation}
		R_{in}^{3}=\frac{N}{2\pi j^{\phi}}=\frac{\displaystyle
		{N^{2}\left(1+\frac{n}{eN}\right)}}{4\pi\Lambda} ~,
	\label{Rpos}
\end{equation}
\end{mathletters}
and impose $R_{in}\geq 0$.
The latter comes to ensure that the outer spacetime would not
develop a conic singularity.

Finally, we are in a position to analyze the conic defect as
a function of the circumferential radius of the tube and the 
total magnetic flux confined.
We do it by studying the interplay of eqs.(\ref{circ},\ref{gamma},
\ref{n/e}), which together determine the conic defect to be
\begin{equation}
	\gamma = \frac{{\displaystyle
	1-(\frac{4n}{eN}-1)G\frac{N^{2}}{\rho^{2}}}}
	{{\displaystyle
	\left(1+G\frac{N^{2}}{\rho^{2}}\right)^{3}}} ~.
\end{equation}

The demand $\gamma\geq 0$ sets a lower bound on the size of
a tube which carries a given amount of magnetic flux
\begin{equation}
	\fbox{$\displaystyle{
	\rho^{2}_{min} = GN^{2}\left(\frac{4n}{eN}-1\right)}$}
	\label{min}
\end{equation}
This establishing one of our main results.
\begin{figure}[tbp]
	\begin{center}
		\includegraphics[scale=0.45]{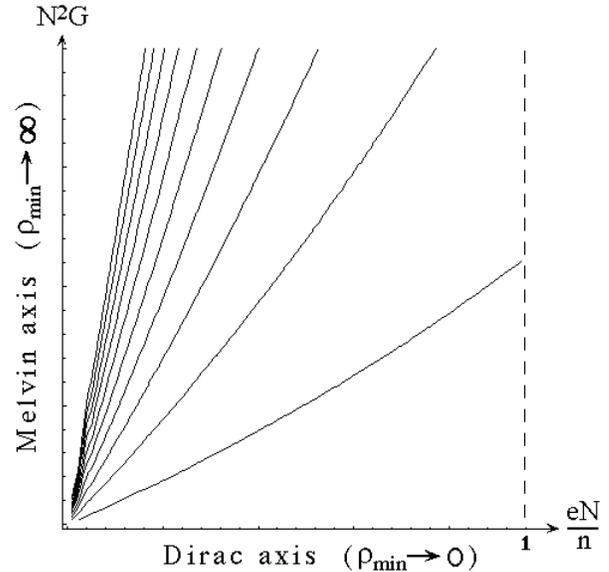}
		\caption{Dirac-Melvin contour plot: Lines of constant
		minimal circumferential radius $\rho_{min}$ range from
		Dirac line $(\rho_{min}\rightarrow 0)$ up to Melvin line
		$(\rho_{min}\rightarrow \infty)$.}
		\label{fig2}
	\end{center}
\end{figure}
Taking into account eq.(\ref{eN/n}), one may further observe
that
\begin{equation}
	\fbox{$\displaystyle{\frac{}{}
	\rho_{min}\geq N\sqrt{3G}}$}
\end{equation}
and is thus driven to a provocative conclusion:
In the presence of gravitation, the Dirac procedure does not
make sense.
Any attempt to arbitrarily squeeze a magnetic flux tube, while
keeping its total magnetic flux fixed, would eventually create
a singularity (and close the surrounding space).
Magnetic flux simply cannot be confined within flux tubes of
sub-Planck cross sections.
\begin{figure}[tbp]
	\begin{center}
		\includegraphics[scale=0.7]{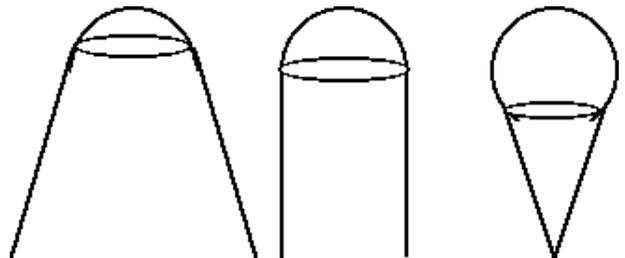}
		\caption{The outer space surrounding a cosmic solenoid is
		conic and open ($\rho\geq\rho_{min}\Rightarrow \gamma\geq 0$)
		or else closed and singular ($\rho <\rho_{min}\Rightarrow
		\gamma<0$).}
		\label{fig3}
	\end{center}
\end{figure}
\subsection{Dual vector field}
In order to decode the Pythagorean quantization condition
eq.(\ref{quan1}), we restrict ourselves to the London limit
of the surface scalar field theory (fixed $\xi= m$).
Starting from
\begin{equation}
	{\cal L}^{S} =
	-m^{2}\gamma^{ab}\Delta_{a}\Delta_{b}-\Lambda ~,
\end{equation}
the $\chi$ equation of motion is
\begin{equation}
	J^{a}_{;a} =\frac{1}{\sqrt{-\det\gamma}}
	\left(\frac{\partial({\cal L}^{S}\sqrt{-\det\gamma})}
	{\partial \chi_{,a}}\right)_{,a}=0 ~.
	\label{chieq}
\end{equation}
The most general solution of this equation is simply
\begin{equation}
	J^{a} = \frac{\varepsilon^{abc}}{\sqrt{-\det\gamma}}
	\partial_{b}q_{c} = 
	\frac{\varepsilon^{abc}}{2\sqrt{-\det\gamma}}f_{bc} ~,
\end{equation}
where $f_{ab} \equiv \partial_{[a}q_{b]}$.

One can perform now a Legendre transformation to exchange the
scalar phase $\chi$ for a surface vector field $q_{a}$.
The prescription then calls for a new Lagrangian
\begin{equation}
	{\cal L}^{S^{*}} = {\cal L}^{S} - J^{a}\chi_{,a}~,
\end{equation}
which up to a total derivative is nothing but
\begin{equation}
	{\cal L}^{S^{*}} =
	-\frac{1}{8m^{2}}\gamma^{ab}\gamma^{cd}f_{ac}f_{bd}
	-e\frac{\varepsilon^{abc}}{\sqrt{-\det\gamma}}
	A_{a}\partial_{b}q_{c}-\Lambda ~.
\end{equation}
In this dual language, $q_{a}$ has been elevated to the level
of a $(2+1)$-dimensional gauge field exhibiting off-diagonal
Chern-Simons interaction\cite{offCS} with the projective
electromagnetic gauge field.
The corresponding equation of motion is given by
\begin{equation}
	f^{ab}_{\hspace{5pt};b} = 2m^{2}e\frac{\varepsilon^{abc}}
	{\sqrt{-\det\gamma}}A_{b,c}~.
\end{equation}

The main question now is the following: What surface gauge 
configuration is equivalent to $\chi=n\phi$?
A simple algebra reveals that the solution consistent with
conditions eqs.(\ref{consist}) is
\begin{equation}
	f_{tz} = -E = const ~.
\end{equation}
It is accompanied by the relations
\begin{eqnarray}
	& \displaystyle{j^{\phi} =
	\frac{eE}{\rho} ~,} & \\
	&\displaystyle{\sigma =
	\frac{eNE}{\rho}+\frac{E^{2}}{2m^{2}} ~.}&
	\label{vec}
\end{eqnarray}
Now, we match the former eq.(\ref{n/e}) with the latter
eq.(\ref{vec}) to obtain 
\begin{equation}
        n - eN = \frac{\rho E}{2m^{2}}~.
\end{equation}

Notice that the configuration in hand is that of a constant
`electric' field in a `ring' capacitor.
It is nothing but the closed $(2+1)$-dimensional analog of the
$(3+1)$-dimensional plate capacitor.
$E$ turns out to be independent of the distance between the
$\pm Q$ `charges', normalized such that
\begin{equation}
	Q \equiv \frac{\rho E}{2m^{2}e} ~.
\end{equation}
The distance between the `charges' can then be taken to
infinity, so that cylindric symmetry is restored without any
lose of generality.
The resulting generalized quantization condition reads
\begin{equation}
	\fbox{$\displaystyle{
	\frac{e\Phi}{2\pi} + \frac{1}{e}Q = n}$}
	\label{newquan}
\end{equation}
It is the combination (magnetic flux) + (dual electric charge),
rather than the magnetic flux by itself, which gets properly
quantized.

The fact that the magnetic flux $N$ does not get quantized comes
with no surprise.
In theories where the complex field extends to spatial infinity
and has non-zero modulus $m$ there, the requirement of finite
energy imposes a strict relation between $A_{\mu}$ and the gradient
of the phase $\chi$ of the complex field.
This relation, in turn, leads to flux quantization because the 
complex field must be single-valued.
The reason that the magnetic flux is not quantized here is that
the complex scalar field, to which the projection of $A_{\mu}$
couples minimally, exists only on the solenoid surface and does
not extend to spatial infinity.

\section{Dilaton Gravity}
Dilaton gravity usually arises as the low energy limit of string
theory, or as the $4$-dimensional effective theory of
higher-dimensional Kaluza-Klein gravity.
The dilaton is a real scalar field that couples to other fields
in a very special way.
Two different metrics are often used in dilaton gravity, related
to each other by means of a conformal transformation.
In the so-called string basis, the dilaton couples to the Ricci
scalar associated with the string metric.
In the so-called Einstein basis, the dilaton couples only to
matter, but does not have a direct gravitational coupling with
the Ricci scalar associated with Einstein metric.
In this context, it is worth mentioning that the extrinsic
curvature surface term prevents the appearance of a dilaton
surface term during the conformal transformation.
This is why we choose to work here in the Einstein basis, where
the self consistent action takes the form
\begin{eqnarray}
	&& Action = \nonumber \vspace{6pt} \\
	&& - \frac{1}{16\pi}\int_{V_{in}}
	(e^{-2k\sqrt G\eta}g^{\mu\lambda}
	g^{\nu\sigma}F_{\mu\nu}F_{\lambda\sigma})^{in}
	\sqrt{-\det g}\,d^{4}x - \nonumber \\
	&& - \frac{1}{8\pi}\int_{V_{in}}
	\left(g^{\mu\nu}\eta_{,\mu}\eta_{,\nu}\right)^{in}
	\sqrt{-\det g}\,d^{4}x + \nonumber \\
	&& + \frac{1}{16\pi G} \int_{V_{in}}{\cal R}^{in}
	\sqrt{-\det g}\,d^{4}x - \nonumber \\
	&& - \frac{1}{8\pi G} \oint_{S} K^{in}
	\sqrt{-\det\gamma}\,d^{3}y + \nonumber \\
	&& + \oint_{S} {\cal L}^{S}\sqrt{-\det\gamma}\,d^{3}y -
	\label{daction} \\
	&& - \frac{1}{8\pi G} \oint_{S} K^{out}
	\sqrt{-\det\gamma}\,d^{3}y + \nonumber \\
	&& + \frac{1}{16\pi G} \int_{V_{out}} {\cal R}^{out}
	\sqrt{-\det g}\,d^{4}x - \nonumber \\
	&& - \frac{1}{8\pi}\int_{V_{out}}
	\left(g^{\mu\nu}\eta_{,\mu}\eta_{,\nu}\right)^{in}
	\sqrt{-\det g}\,d^{4}x + \nonumber \\
	&& - \frac{1}{16\pi}\int_{V_{out}}(e^{-2k\sqrt G\eta}
	g^{\mu\lambda}g^{\nu\sigma}
	F_{\mu\nu}F_{\lambda\sigma})^{out}\sqrt{-\det g}\,d^{4}x ~.
	\nonumber
\end{eqnarray}
The parameter $k$ is recognized as the dilaton coupling constant;
its value depends, however, on the underlying parent theory.
It is normalized such that $k=1$ is dictated by string theory,
whereas $5$-dimensional Kaluza-Klein theory\cite{sqrt3} suggests
$k=\sqrt{3}$.
The canonical fields are now $g_{\mu\nu}(x)$, $A_{\mu}(x)$, the
dilaton $\eta(x)$, and some surface fields as well.
We have some idea how do surface fields serve as electromagnetic
sources, but the dilaton coupling to the surface fields is still
an open question.

\medskip
Now, in analogy with the previous discussion,

\smallskip
\noindent $\bullet$ The variation of the action with respect
to $g_{\mu\nu}(x)$ gives rise to the Einstein equations in the
entire spacetime
\begin{equation}
	\left.{\cal R}_{\mu\nu} - \frac{1}{2}g_{\mu\nu}{\cal R} -
	8\pi G T_{\mu\nu}\right|_{in/out}=0 ~,
	\label{DEin}
\end{equation}
and produces the Gauss-Codazzi matching condition on the surface
\begin{equation}
	\left. {\cal K}_{ab} \right|_{in} +
	\left.{\cal K}_{ab}\right|_{out} = 8\pi GS_{ab} ~,
	\label{DGC}
\end{equation}
only with dilaton modified $T_{\mu\nu}$ and $S_{ab}$.

\smallskip
\noindent $\bullet$ The variation of the action with respect to
$A_{\mu}(x)$ gives the generalized Maxwell equations in the entire
spacetime,
\begin{equation}
	\left.{\left( e^{-2k\sqrt G\eta}F^{\mu\nu}\right)_{;\nu}}
	\right|_{in/out} = 0 ~,
	\label{DMax}
\end{equation}
and the junction conditions on the surface modified to include
the dilaton factor
\begin{equation}
	\begin{array}{ccl}
		& & 4\pi \gamma_{ab}j^{b} = \vspace{6pt} \\
		& & = \left. e^{-2k\sqrt G\eta}F_{\mu\nu}n^{\mu}x^{\nu}_{,a}
		\right|_{in} + \left. e^{-2k\sqrt G\eta}F_{\mu\nu}n^{\mu}
		x^{\nu}_{,a}\right|_{out} ~.
	\end{array}
\end{equation}

\noindent $\bullet$ The variation of the action with respect to
the dilaton field $\eta(x)$ leads to
\begin{equation}
	\left. g^{\mu\nu}\eta_{;\mu\nu} + 
	\frac{1}{2}k\sqrt G e^{-2k\sqrt G\eta}F^{2}
	\right|_{in/out} = 0 ~,
	\label{DILin}
\end{equation}
and the associated dilaton matching condition
\begin{equation}
	\left. n^{\mu}\eta_{;\mu}\right|_{in}+
	\left. n^{\mu}\eta_{;\mu}\right|_{out} = 4\pi\varrho ~.
	\label{DILs}
\end{equation}
The dilaton surface charge $\varrho$ is defined by
\begin{equation}
	\varrho = \frac{\partial{\cal L}^{S}}{\partial\eta}-
	\left(\frac{\partial{\cal L}^{S}}{\partial \eta_{;\mu}}
	\right)_{;\mu} ~.
	\label{varrho}
\end{equation}

\medskip
Given the same symmetry constraints, the above equations admit
the following solution:

\medskip
\noindent $\bullet$ The \textit{inner} solution generalizes
Melvin universe in a straight forward way
\begin{mathletters}
\label{DMelvin}
\begin{equation}
	ds^{2}_{in} = D^{2}(r)(-dt^{2}+dz^{2}+dr^{2}) +
	\frac{r^{2}}{D^{2}(r)}d\phi^{2} ~,
\end{equation}
where
\begin{equation}
	D(r) = M(r)^{\frac{1}{1+k^{2}}} =
	\left(1+\frac{1}{4}GB^{2}r^{2}\right)^{\frac{1}{1+k^{2}}} ~.
\end{equation}
It is accompanied by
\begin{equation}
	F_{r\phi}^{in} = \frac{e^{k\sqrt G\eta_{0}}}
	{\sqrt{1+k^{2}}}\frac{Br}{M^{2}(r)} ~,
\end{equation}
\begin{equation}
	A_{\phi}^{in} = \frac{e^{k\sqrt G\eta_{0}}}
	{\sqrt{1+k^{2}}}\frac{Br^{2}}{2M(r)} ~,
\end{equation}
and the dilaton configuration
\begin{equation}
	\eta^{in} = \eta_{0}-\frac{k}{\sqrt G}\ln{D(r)} ~.
\end{equation}
\end{mathletters}
In accord with $A^{\phi}(0)=0$, it seems reasonable to insist
on $\eta_{0}=0$ as well.
 
\medskip
\noindent $\bullet$ The \textit{outer} solution has again
a flat conic structure
\begin{mathletters}
\label{Dconic}
	\begin{equation}
		ds^{2}_{out} = -dt^{2}+dz^{2}+dr^{2}+
		\gamma^{2}r^{2}d\phi^{2} ~,
	\end{equation}
	\begin{equation}
		F_{r\phi}^{out} = 0 ~,
	\end{equation}
	\begin{equation}
		A_{\phi}^{out} = N ~,
	\end{equation}
but is furthermore characterized by the constant
	\begin{equation}
		\eta^{out} \equiv \eta(r\rightarrow\infty)
		=\eta_{\infty} ~.
	\end{equation}
\end{mathletters}

Substituting eqs.(\ref{DMelvin},\ref{Dconic}) in
eqs.(\ref{DEin}-\ref{varrho}) and keeping the $g_{\mu\nu}(x)$,
$A_{\mu}$, and $\eta(x)$ continuous over the surface, we obtain
a bunch of matching conditions:
\begin{mathletters}
\label{Dmatching}
\begin{equation}
	\lambda = \left(1+\frac{1}{4}GB^{2}R_{in}^{2}
	\right)^{\frac{1}{1+k^{2}}} ~,
\end{equation}
\begin{equation}
	\gamma = \frac{R_{in}}{R_{out}\left(1+\frac{1}{4}
	GB^{2}R_{in}^{2}\right)^{\frac{1}{1+k^{2}}}} ~,
\end{equation}
\begin{equation}
	S^{t}_{t} = S^{z}_{z} =\frac{1}{8\pi G}
	\left(\frac{1}{R_{out}}-\frac{1}{R_{in}
	\left(1+\frac{1}{4}GB^{2}R_{in}^{2}
	\right)^{\frac{1}{1+k^{2}}}} \right) ~,
\end{equation}
\begin{equation}
	8\pi S^{\phi}_{\phi} = -\frac{1}{1+k^{2}}
	\frac{B^{2}R_{in}}
	{\left(1+\frac{1}{4}GB^{2}R_{in}^{2}
	\right)^{\frac{2+k^{2}}{1+k^{2}}}} ~,
\end{equation}
\begin{equation}
	N = \frac{1}{\sqrt{1+k^{2}}}\frac{BR_{in}^{2}}
	{2\left(1+\frac{1}{4}GB^{2}R_{in}^{2}\right)} ~,
	\label{DN}
\end{equation}
\begin{equation}
	4\pi j^{\phi} = \frac{1}{\sqrt{1+k^{2}}}
	\frac{B}{R_{in}\left(1+\frac{1}{4}GB^{2}R_{in}^{2}
	\right)^{\frac{1}{1+k^{2}}}} ~,
	\label{Dj}
\end{equation}
\begin{equation}
	-\frac{k}{1+k^{2}}\ln{\left(1+\frac{1}{4}GB^{2}
	R_{in}^{2}\right)} = \sqrt G\eta_{\infty} ~,
	\label{Deta}
\end{equation}
\begin{equation}
	4\pi\varrho = -\frac{k}{(1+k^{2})}\frac{\sqrt G B^{2}R_{in}}
	{2\left(1+\frac{1}{4}GB^{2}R_{in}^{2}\right)^{\frac{2+k^{2}}
	{1+k^{2}}}} ~.
	\label{Dvarrho}
\end{equation}
\end{mathletters}

The surface fields are subject to the consistency
conditions:
\begin{mathletters}
\label{Dconsist}
\begin{equation}
	S^{t}_{t} = S^{z}_{z} = -\sigma ~,
\end{equation}
\begin{equation}
	S^{\phi}_{\phi}= -Nj^{\phi} =
	\frac{\varrho}{k\sqrt G} ~.
	\label{Dtension}
\end{equation}
\end{mathletters}

Altogether, we can calculate the relevant physical quantities,
the conic defect
\begin{equation}
	\gamma =\frac{1-8\pi G\sigma R_{in}
	\left(1+\frac{1}{4}GB^{2}R_{in}^{2}
	\right)^{\frac{1}{1+k^{2}}}}
	{\left(1+\frac{1}{4}GB^{2}R_{in}^{2}
	\right)^{\frac{2}{1+k^{2}}} } ~,
	\label{Dgamma}
\end{equation}
and the circumferential radius of the flux tube
\begin{equation}
	\rho = \frac{R_{in}}{\left(1+\frac{1}{4}GB^{2}R_{in}^{2}
	\right)^{\frac{1}{1+k^{2}}}} ~.
	\label{Dcirc}
\end{equation}

Eq.(\ref{Dvarrho}) tells us that the dilaton surface charge must
be different from $0$, therefore the surface fields must couple
to the dilaton in some way.
The simplest way to do so is to add a dilaton factor in the 
following way
\begin{equation}
	{\cal L}^{S}= -e^{l\sqrt G\eta}
	\left( \gamma^{ab}(D_{a}\Phi)^{\dagger}(D_{b}\Phi) +
	V(\Phi^{\dagger}\Phi)\right) ~.
\end{equation}
The coupling constant $l$ depends on the underlying theory, the
character of the field, and the dimension of the surface.
Tracing our steps from eq.(\ref{scalar}) to eq.(\ref{eN/n}), we
are led to the following relations
\begin{eqnarray}
	j^{\phi} &=& e^{l\sqrt G\eta_{\infty}}
	\frac{2e\Lambda}{n+eN} ~, \nonumber \\
	\sigma &=& \frac{n}{e}j^{\phi} ~,
	\label{Dn/e} \\
	\varrho &=& -l\sqrt G\sigma ~, \nonumber
\end{eqnarray}
and the subsequent consistency condition
\begin{equation}
	\frac{n}{eN}=\frac{k}{l} \geq 1 ~.
\end{equation}

Giving both $N$ and $\eta_{\infty}$ seems a bit too much, as the
circumferential radius $\rho$ and consequently the conic defect
$\gamma$ get fixed.
Our current interest, however, is to study $\gamma(N,\rho)$ while
holding only $N$ fixed.
This way, for each value of $\rho$, in particular $\rho_{min}$,
one can calculate the attached $\eta_{\infty}$.
Our goal now is to derive the $\rho_{min}$ formula based on the 
physical requirement $\gamma \geq 0$.
We proceed in steps:

\smallskip
\noindent $\bullet$ Start by substituting the relation
$\displaystyle{j^{\phi} = \frac{e}{n}\sigma}$ into eq.(\ref{Dj})
to obtain
\begin{equation}
	4\pi\sigma R_{in}\left(1+\frac{1}{4}GB^{2}R_{in}^{2}
	\right)^{\frac{1}{1+k^{2}}}=\frac{nB}{e\sqrt{1+k^{2}}} ~.
\end{equation}
This enters nicely into the $\gamma=0$ critical condition which
now reads
\begin{equation}
	1 = G\frac{2nB}{e\sqrt{1+k^{2}}} ~.
\end{equation}

\smallskip
\noindent $\bullet$ Next, solve eqs.(\ref{DN},\ref{Dcirc}) to
obtain an expression for $B$
\begin{equation}
	B = \frac{2N\sqrt{1+k^{2}}}{\rho^{2}}
	\left(\frac{R_{in}}{\rho}\right)^{k^{2}-1} ~,
\end{equation}
which can be used to rewrite the critical condition as
\begin{mathletters}
\begin{equation}
	1 = G\frac{4nN}{e\rho^{2}}
	\left(\frac{R_{in}}{\rho}\right)^{k^{2}-1} ~.
\end{equation}
To finish the calculation we only need $\displaystyle{\frac{R_{in}}
{\rho}}$ as a function of $\rho$ (and the fixed $N$).
But such a relation is precisely what eqs.(\ref{DN},\ref{Dcirc})
are capable of producing, namely
\begin{equation}
	\left(\frac{R_{in}}{\rho}\right)^{1+k^{2}} =
	1 + GN^{2}(1+k^{2})\frac{1}{\rho^{2}}
	\left(\frac{R_{in}}{\rho}\right)^{2k^{2}} ~.
\end{equation}
\end{mathletters}

\smallskip
\noindent $\bullet$ The interplay of the two last equations
leads us finally to our main result
\begin{equation}
	\fbox{$\displaystyle{
	\rho^{2}_{min} = G\frac{4nN}{e}
	\left(1-\frac{eN}{4n}(1+k^{2})\right)
	^{\frac{1-k^{2}}{1+k^{2}}}}$}
	\label{Dmin}
\end{equation}
An immediate test of this formula is the recovery of eq.(\ref{min})
at the $k\rightarrow 0$ limit.

Notice that if the dilaton coupling is such that
\begin{equation}
	k^{2}> \frac{4n}{eN}-1> 3 ~,
\end{equation}
there is no lower bound on $\rho$.
It is worth mentioning that the class of $(4+D)$-dimensional
Kaluza-Klein theories, where the extra dimensions form a
$D$-dimensional sphere, induce a $4$-dimensional dilaton gravity
characterized by
\begin{equation}
	k = \sqrt{1+\frac{2}{D}} ~.
\end{equation}
The maximal value of $k$ happens to be precisely $\sqrt{3}$
(achieved for $D=1$).
Thus, in this family of dilaton gravity theories there is
always a bound on the circumferential radius of the solenoid.

A final remark is in order.
The case $k=1$ is singled out by string theory, and as such
deserves special attention.
However, in the context of the present work, it does not seem
to play any exclusive role in the cosmic solenoid game.
One may observe though that the formulae get somewhat simplified,
in particular
\begin{equation}
	\left.\rho_{min}^{2}\right|_{k=1} = GN^{2}\frac{4n}{eN} ~.
\end{equation}

\section{Conclusions}
The Dirac procedure of arbitrarily squeezing a magnetic flux
tube, while keeping its total flux fixed, is conceptually and
substantially modified in the presence of gravity.
A given amount of magnetic flux cannot be confined within a
Planck-scale cross-section tube without closing the surrounding
space and thereby creating a conic singularity.
A similar conclusion holds for dilaton gravity (and string theory)
as well provided the  dilaton coupling is sub-critical.

In this paper, we have constructed a self-consistent general
relativistic configuration which describes such a finite
cross-section magnetic flux tube.
The so-called cosmic solenoid is modeled by an elastic
superconductive surface which separates the inner Melvin core
from the surrounding flat conic geometry.
The Gauss-Codazzi (and electromagnetic) matching conditions are
derived by means of a self-consistent action where the source term,
which governs the surface currents, is sandwiched between internal
and external gravitational surface terms.
Surface superconductivity is realized by means of either a complex
Higgs scalar minimally coupled to projective electromagnetism, or
alternatively by a dual surface gauge field with off-diagonal
Chern-Simons interaction with projective electromagnetism.
It is surface field theory which dictates the vital connection
between surface current and surface energy density, which in turn
links the conic defect to the circumferential radius.
Our analytic analysis produces a generalized quantization condition, 
namely
$$
(magnetic\,\,flux) + (dual\,\,electric\,\,charge) = integer ~,
$$
thereby allowing for a non-trivial Aharonov-Bohm effect.

\acknowledgments
It is our pleasure to thank Prof. Yakir Aharonov for the valuable 
discussions and enlightening remarks.

\end{document}